\documentstyle[prl,aps,floats,epsf,psfig,colordvi]{revtex}
\draft
\def\np#1#2#3   {{ Nucl. Phys.} {\bf#1}, #2 (#3). }
\def\pcps#1#2#3 {{ Proc. Cam. Phil. Soc.} {\bf#1}, #2 (#3). }
\def\pl#1#2#3   {{ Phys. Lett.} {\bf#1}, #2 (#3). }
\def\plc#1#2#3   {{ Phys. Lett.} {\bf#1}, #2 (#3); }
\def\prep#1#2#3 {{ Phys. Rep.} {\bf#1}, #2 (#3). }
\def\prev#1#2#3 {{ Phys. Rev.} {\bf#1}, #2 (#3). }
\def\prl#1#2#3  {{ Phys. Rev. Lett.} {\bf#1}, #2 (#3). }
\def\prs#1#2#3  {{ Proc. Roy. Soc.} {\bf#1}, #2 (#3). }
\def\ptp#1#2#3  {{ Prog. Th. Phys.} {\bf#1}, #2 (#3). }
\def\rmp#1#2#3  {{ Rev. Mod. Phys.} {\bf#1}, #2 (#3). }
\def\rpp#1#2#3  {{ Rep. Prog. Phys.} {\bf#1}, #2 (#3). }
\def\zp#1#2#3   {{ Z. Phys.} {\bf#1}, #2 (#3). }
\def\epj#1#2#3   {{ Eur. Phys. Jour.} {\bf#1}, #2 (#3). }

\begin{document}

\wideabs{
\title{Reply to comment on parton distributions, $d/u$, and higher twist effects
at high $x$ }
\author{ U.~K.~Yang and A.~Bodek}
\address{Department of Physics and astronomy,
University of Rochester, Rochester, NY 14627 }

\maketitle
\twocolumn
}

M. Melnitchouk {\it et al.}~\cite{comment} misunderstand the
model of Frankfurt and Strikman~\cite{FS} for nuclear binding
effects in deep inelastic scattering (DIS). In addition,
their comment is entirely irrelevant to
the results of our article.

In the Frankfurt-Strikman model the
main parameter  used to describe
the deviations of the structure functions
of bound nucleons from those of free nucleons
is the average kinetic energy of the nucleons in the nucleus, with small
corrections due to energy binding (the parameter is 
$k^2/2m +\epsilon_A$, where $k$ is nucleon momentum and $\epsilon_A$
is energy binding per nucleon ($\epsilon_{A\sim 200} \approx 8 MeV$).
If the value of $x$ is not too large
(below $x=0.75$)  the binding effects are proportional to the average
value  $\left<k^2\right>/2m +\epsilon_A$.
For heavy nuclei one can safely approximate the 
resulting A-dependence of the binding effects
in terms of the A- dependence  of $\left<k^2\right>/2m$.
Note that the Fermi motion effects are
also proportional to average value of $k^2$ in this x-range.
%
%
Frankfurt and Strikman also argue \cite{FS} that a similar
pattern is valid in a wide range of models including pion models and 
the so called nuclear binding models where
the large x depletion of the nuclear structure functions is from a
reduction of the light-cone momentum fraction carried by nucleons.
Hence the overall deviation from 1 of the ratio of the nuclear
structure functions to that of free nucleons ($R_A(x)-1$) is 
a factorized function of $\phi(A)*f(x,Q^2)$.
An estimate of the overall scale of $\phi(A)$
and the function $f(x,Q^2)$ can be extracted from
the SLAC data on heavy targets. It is well known that in mean field
nuclear models, $\left<k^2\right>/2m$ is proportional to the average nuclear density 
(this is also approximately valid for $\left<k^2\right>/2m+ \epsilon_A$).
The nuclear density fit gives a good description of the SLAC data~\cite{SLAC}.
%
%

We have used the SLAC nuclear density fit to get the
effects of the nucleon binding in the deuteron
on the structure functions.
The SLAC  fit implies that the effects in the deuteron
are about 25\% of the effects in iron. For comparison,
Frankfurt and Strikman~\cite{FS} calculate the following
ratio for the
relevant quantity in iron and deuterium
${\left<k^2\right>_{Fe} +\epsilon_{Fe}\over \left<k^2\right>_D+\epsilon_D}\approx 5$. 
From this ratio, they extract
the relation  $(F_{2D}/F_{2N}-1)= 0.25(F_{2Fe}/F_{2D}-1) $,
which is close to the value used in our paper. 
Therefore, although the notion
of nuclear density for the deuteron may not be very well defined,
the value of the nuclear density for deuterium that was used in the SLAC fit
yields a similar correction for nuclear
binding in the deuteron as the 
estimate by Frankfurt and Strikman.
Note that the the uncertainty in the value of $0.25$ is about $20\%$, and 
that the x-dependence is well constrained by the $A\ge 4$ data.

Figure 1 of our paper shows that the nuclear
binding effects in deuterium as estimated
from the nuclear density (solid line) are actually
almost identical to the model of Melnitchouk and Thomas~\cite{MT}
(dashed line) in the region of $x = 0.6$.  Therefore, all
three models for
for the nuclear binding corrections to the structure functions
in the deuteron, namely
the SLAC nuclear density fit, the Frankfurt and Strikman model, and the
Melnitchouk and Thomas 
model are all in agreement with the corrections that are
used in our paper at large $x$.
The proportionality of the nuclear binding corrections
to the average kinetic energy $k^2$ is pretty generic and
hence holds in the model of Melnitchouk and Thomas
 as well. 
When one averages over nucleon momenta one ends up with a similar
combination of kinetic energy and energy binding for their
model also.

However, the predictions from the Melnitchouk and Thomas model
(which so far has not been applied to $A\ge 4$  nuclei)
show large binding effects in deuterium at $x = 0.3$, where there is
no difference between the structure functions of iron and deuterium.
At $x = 0.2$, the model predicts binding effects
in deuterium which are of opposite sign to that in iron. These
strange features of the Melnitchouk-Thomas model
are the reasons why we did not use that model in the extraction of
neutron structure functions from deuterium data.
The strange features of their model at lower x are likely to
originate from the fact the  energy momentum sum rule is not satisfied
in their theory. A resolution of this problem and tests of the model
for heavy nuclei are needed. 

If light nuclei like $^3$He, $^3$H are considered,
the original formulae of Frankfurt and  Strikman
in terms of $k$ and $\epsilon$ (with realistic models of the A=3 wave functions)
should be used. Here also, there is great advantage
of relating the nuclear binding effects in a light nucleus to
the experimental ratio of the binding effects in iron and deuterium.
Within such a framework, the energy momentum sum rules are satisfied,
and the models can be used over a larger range of x.

\end{document}